# Disentangling stellar atmospheric parameters in astronomical spectra using Generative Adversarial Neural Networks.

## Application to Gaia/RVS parameterization.

M. Manteiga[1, 2], R. Santoveña[2, 3], M.A. Álvarez[2, 3], C. Dafonte[2, 3], M.G. Penedo[2, 3], S. Navarro[4], and L. Corral[4]

[1] Universidade da Coruña (UDC), Department of Nautical Sciences and Marine Engineering, Paseo de Ronda 51, 15011, A Coruña, Spain
e-mail: manteiga@udc.es
[2] CIGUS CITIC, Centre for Information and Communications Technologies Research, Universidade da Coruña, Campus de Elviña s/n, 15071 A Coruña, Spain
[3] Universidade da Coruña (UDC), Department of Computer Science and Information Technologies, Campus Elviña s/n, 15071 A Coruña, Spain
[4] Universidad de Guadalajara, Instituto de Astronomía y Meteorología, Jalisco, México

**ABSTRACT**

*Context.* The rapid expansion of large-scale spectroscopic surveys has highlighted the need to use automatic methods to extract information about the properties of stars with the greatest efficiency and accuracy and also to optimise the use of computational resources.
*Aims.* A method based on Generative Adversarial Networks (GANs) is developed for disentangling the physical (effective temperature and gravity) and chemical (metallicity, overabundance of α-elements with respect to iron) atmospheric properties in astronomical spectra. Using a projection of the stellar spectra, commonly called latent space, in which the contribution due to one or several main stellar physicochemical properties is minimised while others are enhanced, it was possible to maximise the information related to certain properties, which can then be extracted using artificial neural networks (ANN) as regressors with higher accuracy than a reference method based on the use of ANN trained with the original spectra.
*Methods.* Our model utilises autoencoders, comprising two artificial neural networks: an encoder and a decoder which transform input data into a low-dimensional representation known as latent space. It also uses discriminators, which are additional neural networks aimed at transforming the traditional autoencoder training into an adversarial approach, to disentangle or reinforce the astrophysical parameters from the latent space. The GANDALF tool is described. It was developed to define, train, and test our GAN model with a web framework to show how the disentangling algorithm works visually. It is open to the community in Github.
*Results.* The performance of our approach for retrieving atmospheric stellar properties from spectra is demonstrated using Gaia Radial Velocity Spectrograph (RVS) data from DR3. We use a data-driven perspective and obtain very competitive values, all within the literature errors, and with the advantage of an important dimensionality reduction of the data to be processed.

## 1. Introduction

The explosive development of massive spectroscopic surveys in recent decades has highlighted the convenience of addressing their exploitation through automatic techniques, often supported by Machine Learning (ML) methods. It is now a fact that ML have become indispensable in astronomy, enabling researchers to handle the vast amounts of data generated by modern telescopes and surveys. Among the examples available in the literature, we can highlight several applications in various fields. In the area of spectroscopic classification, the use of neural networks and support vector machines two decades ago significantly improved the accuracy and efficiency of classifying stars, galaxies, and quasars based on their spectral features. Examples can be found in the pioneering works by Bailer-Jones et al. (1998) and Re Fiorentin et al. (2007) based on the Sloan Digital Sky Survey (SDSS, Abazajian et al. (2009)). More recent works, such as those focused on the Gaia survey (Gaia Collaboration et al. 2016), use Gaussian mixture models and Decision Trees as in Hughes et al. (2022) and Delchambre et al. (2023).

In the problem of classification outlier analysis, self-supervised learning techniques like autoencoders and clustering methods such as self-organised maps have been used to detect unusual spectral signatures, which can indicate weird or previously unknown astronomical objects (Baron & Poznanski (2017); Dafonte et al. (2018)). ML has also been applied to detect exoplanets in light curves from missions like Kepler (Barbara et al. 2022) and TESS (Tardugno Poleo et al. 2024). Deep learning models, such as Convolutional Neural Networks (CNN), have been particularly successful in identifying the subtle dips in brightness that indicate a planet transiting its host star (Shallue & Vanderburg 2018) or, more recently, in detecting hidden molecular signatures in cross-correlated spectra from exoplanetary targets (Garvin et al. 2024), to mention a couple of examples.

One of the most effective methods in ML for improving knowledge extraction from large datasets is using Representation Learning (RL) algorithms. One of the capabilities of RL is to transform a high-dimensional input vector into a lower-dimensional vector representation. After suitable training, these representations enable models like t-SNE or UMAP to identify patterns or anomalies within the data. While RL is a broad concept encompassing all tevvvchniques to learn meaningful data representations, Disentangled Representation is a learning approach in which ML models are designed to acquire representations that can identify and separate the underlying factors embedded in the observed data. Disentangled representation learning (DRL) helps in producing explainable representations of the data, where each component has a clear semantic meaning (Bengio et al. 2013; Lake et al. 2017).

One approach in Representation Learning is Generative Modelling (Bishop 2006) which also consists of a self-supervised learning model to construct a vector representation of the input data. As a result, the model can be used to generate new data from the learned distribution. In particular, Generative Adversarial Networks (GANs, Goodfellow et al. 2014) address the problem with two submodels: the generator model which is trained to generate new examples, and the discriminator model that tries to classify the examples as real (from the domain) or synthetic (generated). The two models are trained simultaneously in a zero-sum, adversarial setting, where the generator aims to produce increasingly realistic samples, while the discriminator seeks to distinguish between real and synthetic data. Training continues until the discriminator is unable to reliably differentiate between the two, indicating that the generator is producing plausible examples. In a generative configuration, the adversarial discriminator can be used to assess the quality of a reconstructed signal for which conditioning factors do not exist in the training set (Mathieu et al. 2016; Szabó et al. 2017; Chen et al. 2016). Another alternative, proposed in Fader Networks (Lample et al. 2018), is to apply the adversarial discriminator on the latent space itself, to prevent it from containing any information about the specified conditioning factor.

There is abundant literature on separating distinct, independent factors or components of information within a complex dataset or representation, in what has been called information disentangling. In the context of ML, disentangling aims to isolate meaningful, interpretable features—such as specific physical parameters or latent factors—so they are not entangled with other, unrelated information. GANs can play a significant role in information disentangling by learning to generate complex data distributions while isolating meaningful, interpretable features within the data. They have been used in many domains, typically involving image treatment or reconstruction. For instance, Rifai et al. (2012) applied generative convolutional networks to facial expression recognition; Cheung et al. (2015) used autoencoders and a cross-covariance penalty to recognise handwritten style and Chen et al. (2016) applied GANs to several image recognition problems. In a recent study by Wang et al. (2024), the current state of the literature is comprehensively reviewed, providing an in-depth discussion of various methodologies, metrics, models, and applications.

In the astrophysical domain, the problem of stellar atmospheric parameter ($T_{eff}$, log$g$, [M/H] and [$\alpha$/Fe]) disentangling applied to stellar spectra was first studied by Price-Jones & Bovy (2019) who fitted a polynomial model of the non-chemical parameters ($T_{eff}$ and log$g$) to synthetic spectra in a grid and then used principal component analysis and a clustering algorithm to identify chemically similar groups.

Disentangling using neural networks has been recently addressed in the works by de Mijolla et al. (2021) de Mijolla & Ness (2022). In de Mijolla et al. (2021), a neural network with a supervised disentanglement loss term was applied to a synthetic APOGEE-like data set of spectra to find chemically identical stars without the explicit use of measured abundances. In a more recent work Santoveña et al. (2024) (Paper 1) developed a method, also based on GANs, for disentangling the physical (effective temperature and gravity) and chemical (metallicity, overabundance of $\alpha$-elements and individual elementary abundances) properties of stars in astronomical spectra. In this case, an original version of adversarial training was developed with the novelty of making use of one discriminator per stellar parameter to be disentangled. The methodology was demonstrated using APOGEE and Gaia/RVS synthetic spectra.

In this article, we probe the use of GANs to address the problem of extracting stellar atmospheric parameters from observational stellar spectra with higher precision in the derived values and higher computational efficiency than the reference method based on the use of ANN trained with the spectra. Additionally, we make available to the astrophysical community our tool, GANDALF (Generative Adversarial Networks for Disentangling and Learning Framework) developed to define, train, and test our GAN model with a web framework to show how the disentangling algorithm works. The paper is structured as follows: Section 2 describes the specific deep-learning architecture of our algorithm with multi-discriminators as well as briefly introduces GANDALF the web-based framework for training, testing, and visualising the disentangled models. Section 2 also describes the use of a dimensionality reduction technique, t-SNE (t-distributed stochastic neighbour embedding van der Maaten & Hinton 2008), to validate our disentangling results. Section 3 presents the application of our algorithm for extracting stellar parameters using disentangling and entangling (in the sense of reinforcement, as will be shown later) in one massive spectroscopic survey, the RVS Gaia spectroscopic survey.

By using a projection of the signal (stellar spectra), commonly called latent space, in which the contribution due to one or several main stellar physicochemical properties ($T_{eff}$, log$g$, [M/H] and [$\alpha$/Fe]) is minimised while others are enhanced, it was possible to maximise the information related to certain properties, that can then be extracted with higher accuracy and computational efficiency.

To test the method we use observations of the Gaia instrument RVS published in DR3 Gaia release (see for instance Recio-Blanco et al. 2023). We rely on 64 305 reference stars to train and test the networks, with accurate measurements of their atmospheric parameters in the literature. The quality of our parameterisation is presented through the statistics against the literature values and using the Kiel ($T_{eff}$ vs. log$g$) and [M/H] vs. [$\alpha$/Fe] diagrams, as well as in comparison with the reference baseline algorithm, ANN trained with the stellar spectra. Finally, section 4 summarises the main results and discusses the advantages that our method and our disentangling framework GANDALF can offer the community within the chosen application domain.

## 2. GANDALF: Generative Adversarial Networks for Disentangling and Learning Framework

### 2.1. Generative Artificial Networks with multidiscriminators

#### 2.1.1. Architecture

As previously mentioned, our GAN model uses two sub-models, a generator and a discriminator. The generator utilises autoencoders, comprising two artificial neural networks: an encoder $E$ that transforms input data $\mathbf{x}$ into a low-dimensional representation known as latent space, denoted as $E(\mathbf{x}) = \mathbf{z}$; and a decoder $D$ that takes this abstract representation as input to reconstruct the original data, $D(\mathbf{z}) = \bar{\mathbf{x}}$. The second sub-model involves employing discriminators, and neural networks aimed at transforming the traditional autoencoder training into an adversarial approach, to disentangle astrophysical parameters from the latent space. In our approach, we adapt the autoencoder architecture to prioritise the elimination of one or several stellar atmospheric parameters (**y**). This modification involves inputting such astrophysical parameters twice:

1. Initially, the encoder is adjusted to accept both the spectrum and the parameter (**y**) as input, denoted as $E(\mathbf{x}, \mathbf{y}) = \mathbf{z}$. This modification to the traditional GANs facilitates encoding the spectrum-related information, aiding in its effective elimination during training.
2. Subsequently, the astrophysical parameter is incorporated into the latent space as input for the decoder, denoted as $D(\mathbf{z}, \mathbf{y}) = \bar{\mathbf{x}}$, to allow it to reconstruct the original data (**x**). If the discriminators work properly, the latent space (**z**) will have no information about the parameter (**y**), and hence the decoder could not reconstruct the spectrum (**x**) exclusively from (**z**).

The autoencoder tries to reconstruct the original input through the encoding and decoding phases. For the computation of errors, we use the classical mean squared error (MSE), calculated as the difference between the reconstruction of the decoder ($\bar{\mathbf{x}}$) and the original spectrum $\mathbf{x}$.

$$loss_{rec} = \frac{1}{n} \sum_{i=1}^{n} (x_i - \bar{x}_i) \tag{1}$$

The discriminator receives as input the latent space $\mathbf{z}$ generated by the encoder and tries to predict the stellar atmospheric parameters (**y**). It is trained to minimise the error when predicting the value of the stellar parameters. The error is computed using the cross-entropy or log loss (logarithmic loss or logistic loss, Cover 2006), which calculates the difference between two probability distributions, the expected probability of (**y**) and the probability of the predicted values ($\bar{\mathbf{y}}$).

$$loss_{DISC} = -\frac{1}{n} \sum_{i=1}^{n} y_i \cdot log(\bar{y}_i) + (1 - y_i) \cdot log(1 - \bar{y}_i) \tag{2}$$

Finally, the autoencoder objective is modified not only to achieve a correct reconstruction of the original spectrum (**x**) but also to try to maximize the errors of the discriminators. To control the trade-off between the two objectives, a parameter $\lambda$ is used in the computation of the autoencoder loss.

$$loss_{AE} = loss_{rec} - \lambda \cdot loss_{DISC} \tag{3}$$

If the multi-discriminator approach is applied, as will be shown later, a parameter $\lambda_i$ can be defined for each discriminator.

While our model draws inspiration from generative models like GANs or cGANs (convolutional GANs), there are distinctions, particularly in the use of discriminators. Unlike traditional methods where the discriminator's input is the generator's output, in our approach, adversarial training targets the latent space, acting as input to the discriminators. The role of the generator is assumed by the decoder, which utilises stellar parameters (**y**) and latent space to generate and reconstruct output data. This architecture is influenced by Fader Networks, proposed by Lample et al. (2018), adapted for image reconstruction based on discrete attributes. We adapt these concepts to an astronomical context, applying the networks to stellar spectra. In this case, images with discrete attributes are replaced by astrophysical parameters with continuous values.

To address the challenges posed by the need to discretise such possible values of the astrophysical parameters, we implement a multi-discriminator approach, assigning one discriminator per parameter to mitigate the combinatorial explosion caused by discretisation. Each discriminator receives the latent space (**z**) as input and produces a vector indicating the probability of belonging to each discretised bin for its corresponding parameter ($p(y^d|z)$) (in Figure 1). For each parameter, the domain is divided into 10 intervals of equal size. This approach significantly reduces computational complexity. For instance, if we process 5 parameters using 10 bins each, the multi-discriminator approach minimises the number of classes from 100,000 to 50, resulting in more manageable training. We validate this improvement through comparison with traditional methods in Section 3.3. More details on the algorithm can be found in Paper 1.

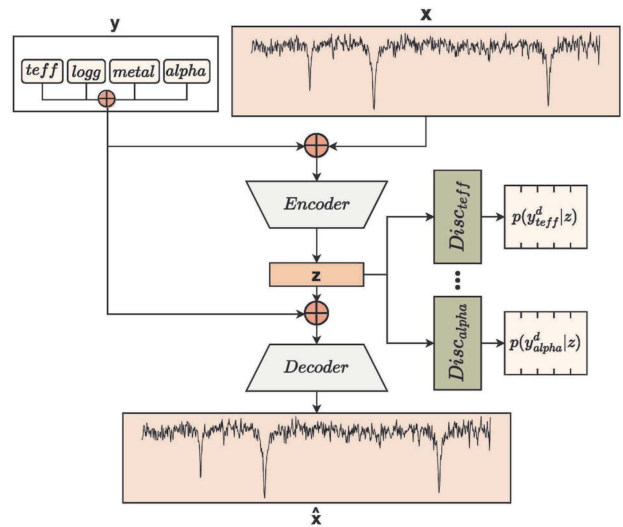

**Fig. 1.** The disentanglement architecture featuring multi-discriminators operates as follows: Spectra $\mathbf{x}$ and stellar parameters $\mathbf{y}$ are concatenated as input. The encoder then maps this concatenated input, $\mathbf{x} \oplus \mathbf{y}$, to the latent space $\mathbf{z}$, aiming to disentangle $\mathbf{y}$ from $\mathbf{z}$. Subsequently, the decoder's task is to reconstruct $\mathbf{x}$ using $\mathbf{z} \oplus \mathbf{y}$ as input. In parallel, discriminators receive $\mathbf{z}$ as input and endeavour to predict the astrophysical parameters within a discretised space $\mathbf{y}^d$.

#### 2.1.2. Implementation

Grid search techniques were employed to determine the model's optimal configuration and identify suitable setups. We analyse the reconstruction errors versus the discriminators' errors, to maintain a balance between them. The size of the latent space was selected by evaluating the relationship between size and the information loss in the reconstruction. Additionally, GAN-DALF was utilised to facilitate swift and parameter adjustments for both the model and the training process. The configuration employed for the results presented in this study is outlined below.

Based on the tests, the encoder, decoder, and discriminators are constructed as fully connected neural networks. The encoder comprises two hidden layers with 512 and 256 neurons, respectively. Its output layer, representing the latent space (**z**), consists of 25 neurons. Conversely, the decoder mirrors the encoder's configuration, featuring two hidden layers with 256 and 512 neurons, respectively. The input size may vary depending on the number of astrophysical parameters to be disentangled, with a base size of $256 + n\_params$. All discriminators share the same structure, with two hidden layers comprising 64 and 32 neurons, respectively. The output size of the discriminators is contingent upon the number of bins utilised for discretisation; in this instance, 10 bins per parameter were employed.

In the multi-discriminator approach, each discriminator $i$ is assigned a $\lambda_i$ value, enabling control over the significance attributed to eliminating each stellar parameter.

$$loss_{AE} = loss_{rec} - \sum_{i=1}^{n_{disc}} \lambda_i \cdot loss_{DISC_i} \qquad (4)$$

An autoencoder small loss term is obtained not only when the autoencoder can reconstruct the original spectrum accurately (small $loss_{rec}$) but also when the discriminator presents large errors (large $loss_{DISC_i}$). The balance between both loss terms is controlled by $\lambda_i$ for each stellar parameter, allowing either disentangling ($\lambda_i > 0$) or reinforcement ($\lambda_i < 0$) of each specific stellar parameter. We use the term entangling for the $\lambda_i < 0$ case, as opposed to disentangling (see section 2.3 for details).

### 2.2. GANDALF framework

Our disentangling framework, GANDALF, comprises several Python classes designed for data generation, a command-line tool facilitating model definition, training, and testing, and a web application providing visual insights into the algorithm's functioning. GANDALF is openly accessible to the community as an open-source tool hosted on GitHub[1]. Figure 2 illustrates our framework, which aims to empower the community to use and tailor our disentangling approach to their respective domains.

### 2.3. GANDALF for entangling

The GANDALF framework was built to facilitate and automate the disentangling architecture described above. Thanks to its flexibility, this architecture can be modified to be adapted to different scenarios, for instance to different spectra size or resolution. Beyond adjusting the architecture or altering the configuration of the networks within its structure, several training parameters can also be modified. Among them, $\lambda_i$, the parameter that controls the trade-off between the loss term in the reconstruction of the spectrum and the loss term of a specific discriminator, has very interesting use cases. When $\lambda_i$ is negative, the effect of the discriminator's error on the autoencoder's loss function is inverted compared to when $\lambda_i$ is positive. This means that for $\lambda_i < 0$ the errors of the discriminators as well as those of the spectrum reconstruction must be minimized to guarantee that the autoencoder loss function is small. As a consequence, the training process tends to reinforce the presence in the latent space of the stellar parameter $i$ related to $\lambda_i$. This "entangling" technique (as opposed to disentangling) can be done with GANDALF using a simple flag. It can also be combined with the multi-discriminator approach since GANDALF allows a $\lambda_i$ factor adjustment for each of the conditional parameters. In this way, following equation 4, positive values favour the elimination of a stellar parameter (disentangling) while negative values preserve their presence (entangling).

---

[1] https://github.com/raul-santovena/gandalf

### 2.4. t-SNE for assessing the level of disentanglement

The t-distributed stochastic neighbour embedding (t-SNE) algorithm is a statistical technique primarily employed to visualise high-dimensional data by projecting them in a lower-dimensional representation (in our case, two dimensions) in a way that preserves local relationships in the data. Specifically, t-SNE arranges similar data points close to one another on a two-dimensional map, thus placing similar spectra in proximity on the map. Since the morphology of a spectrum is influenced by stellar atmospheric parameters, highlighting the distribution of a specific stellar parameter on the t-SNE map provides a visual indication of the correlation between the shape of the spectrum and that parameter. This allows us to leverage the visual effectiveness of t-SNE to offer validation support for our disentangling method. If the disentangling method effectively separates stellar parameters from other spectral information, the distribution seen after applying t-SNE to latent space representations should show no correlation, or at least a reduced correlation, with those parameters compared to the original data representation. This means that when we colour-code the objects according to the disentangled parameter in a t-SNE representation, we will obtain a random distribution of colours, indicative that such a parameter is not present in the current latent representation of our objects. On the contrary, those stellar parameters that were not disentangled (or that even were entangled) will condition the shape of the latent representation and if we tagged their values in the t-SNE diagram we will see an ordered distribution. In section 3.3, we use the t-SNE algorithm to validate the level of disentanglement obtained through generative networks using GANDALF.

## 3. Application of GANs to Gaia/RVS parameterisation

In June 2022, Data Release 3 (DR3) was unveiled, comprising approximately 1 million medium-resolution spectra from the Gaia satellite's Radial Velocity Spectrometer (RVS, Recio-Blanco et al. 2023). This dataset also includes estimations of the main stellar parameters derived from these spectra. The RVS instrument generates medium-resolution spectra with $R = 11,500$ in the near-infrared electromagnetic spectral region. These spectra cover a wavelength range spanning from 8460 to 8700 Angstroms (Å), focusing specifically around the lines of the Calcium triplet. The analysis of RVS spectra in DR3 was carried out by DPAC, the international consortium responsible for processing Gaia mission data, which extracts radial velocities and astrophysical parameters from the observations. This estimation of parameters, which we refer to as stellar parameterisation, involves a model-driven approach, where the observed spectra are interpreted through comparison with a grid of synthetic spectra (Recio-Blanco et al. 2023).

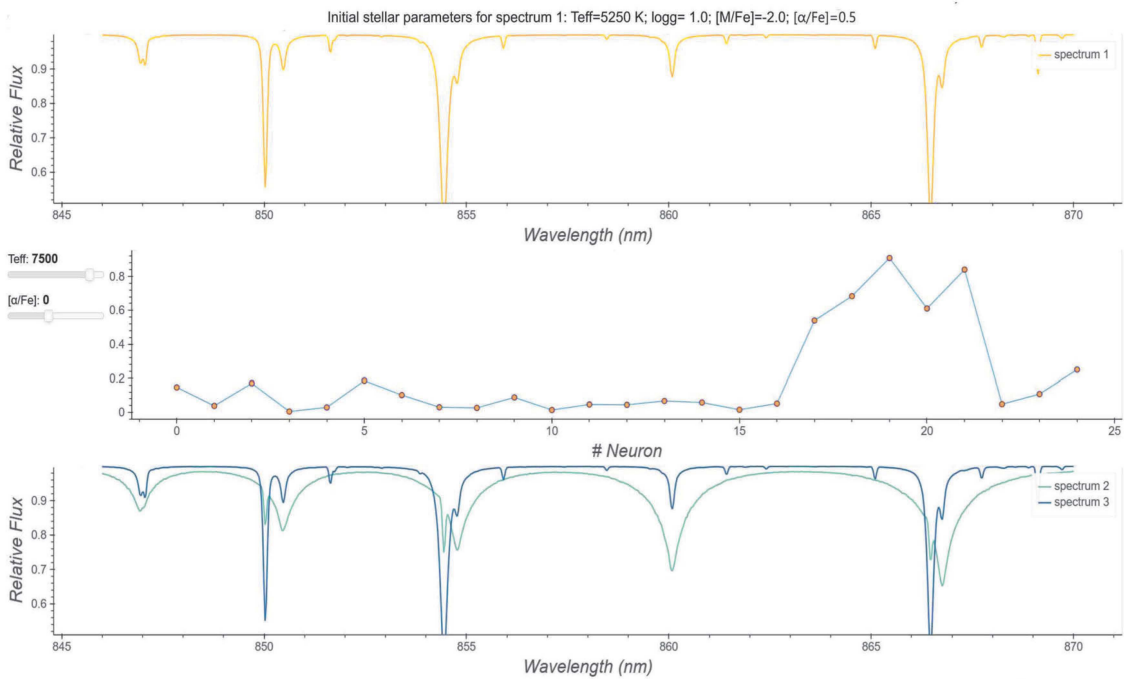

**Fig. 2.** We show how GANDALF, once trained, can generate modified versions of an initial spectrum by altering the stellar parameters within the latent space. The top panel displays the original spectrum along with its four stellar parameters. In the middle panel, the latent space representation is shown. In the bottom panel, two spectra are displayed: a new spectrum generated by setting the stellar parameters to $T_{eff} = 7500$ K and $[\alpha/Fe] = 0.0$ (spectrum 2), and the original spectrum reconstructed by the autoencoder (spectrum 3), which is nearly identical to the initial input (spectrum 1). Figure adapted from GANDALF application.

In this work, we propose a different approach which consists of stellar parameterisation using neural networks trained not with models but with data (spectra) from a set of reference stars with a reliable determination of stellar atmospheric parameters in the literature. In this data-driven context, we want to test the use of GAN with or without disentangling and entangling, to optimise the parameterisation process.

### 3.1. Reference stars for atmospheric stellar parameterisation

To train and test our networks, we used a reference catalogue of stars with atmospheric parameters in two large high-resolution spectroscopic surveys: APOGEE DR17 (Majewski et al. 2017; Wilson et al. 2019; Abduro'uf et al. 2022), and GALAH DR3 (Buder et al. (2021)). Only stars in the RVS magnitude interval $8 \leq G \leq 14$ were considered. We removed some problematic stars in both samples, those with STAR_BAD or FE_H_FLAG flagged from the APOGEE sample and those with flag_sp or flag_fe_h values set to 0 from the GALAH sample. The sample consisted of 64305 stars, most from APOGEE DR17 (54 422) complemented with 9883 from GALAH DR3. According to the parameterisation pipelines, the errors in the parameters are the following: $T_{eff} \leq 200K$; $\log g \leq 0.5$ dex; [M/H] $\leq 0.3$ dex and $[\alpha/Fe] \leq 0.2$ dex. Note that APOGEE stops at -2.5 dex which is up to where the ASPCAP pipeline (García Pérez et al. 2016; Holtzman et al. 2018) assigns metallicities. Figure 3 shows the distribution of the four main atmospheric parameters for the reference sample of stars. Our dataset is centred on FGK giant and dwarf stars with chemistry similar to that of the Sun. Still, it includes also a small sample of low metallicity stars, and some stars with $[\alpha/Fe]$ enhancement, with values around 0.3 dex. In Figure 4 and 5 we present the Kiel diagram as well as [M/H] vs. $[\alpha/Fe]$ diagram for the complete dataset of reference stars.

### 3.2. GANDALF and DR3

GANDALF framework is applied to create a new representation of RVS spectra for two main objectives. First, we aim to improve the accuracy of parameterization obtained by ANN trained as regressors by using the generated latent space as input. Second, this approach seeks to enhance inference times for parameterisation, benefiting from the more compact size of the new representation of the spectra.

To extract information from a set of observational spectra for stars with unknown atmospheric parameters, it is essential to adapt the architecture described before and shown in Figure 1, initially designed for synthetic spectra. In the case of synthetic spectra stellar parameters are introduced at the input along with the spectra. Figure 6 exemplifies the change in architecture. In this case, the stellar parameters are only concatenated with the latent space as an input for the decoder during the training phase. This architecture loses effectiveness compared to the original one but it is also much less restrictive.

As mentioned earlier, during the training phase with observed stellar spectra, the architecture in Figure 6 is applied to a dataset with known astrophysical parameters. A model is then developed to either eliminate or preserve the conditional parameters. In the inference phase, the trained model predicts parameters for new, unseen spectra, where the stellar parameters are unknown. The schematic for this phase is shown in Figure 7. Once the encoder is trained to generate a latent space, it can process new spectra to produce their representations in that space.

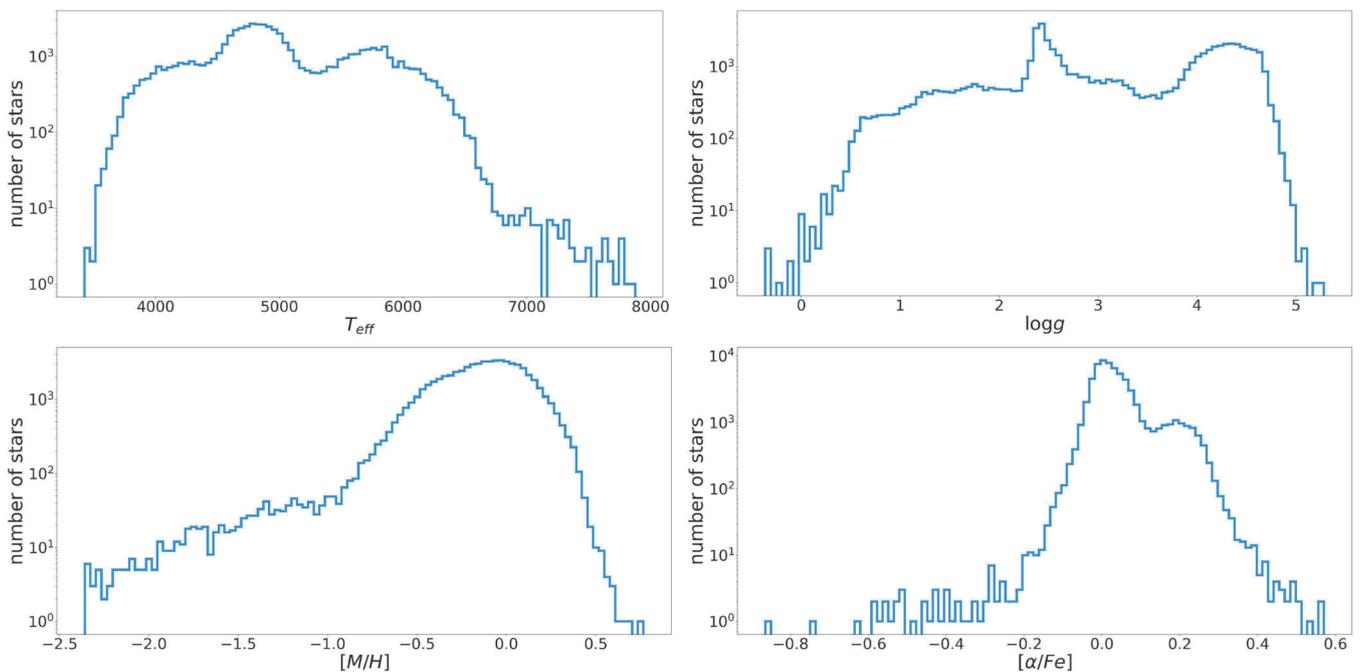

**Fig. 3.** Distribution of the four main stellar atmospheric parameters in our sample of stars. Values from APOGEE DR17 and GALAH DR3.

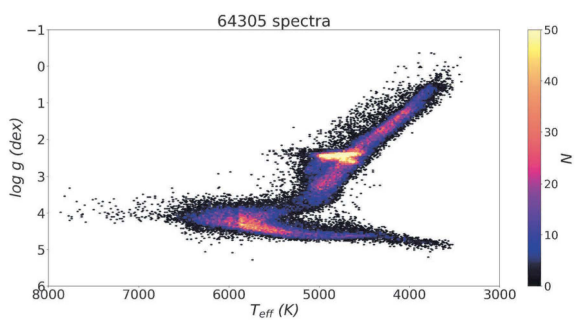

**Fig. 4.** Kiel diagram showing the gravity ($\log g$) against the effective temperature ($T_{eff}$) for the selected sample of reference stars, with parameter values obtained from APOGEE DR17 and GALAH DR3, see text for details.

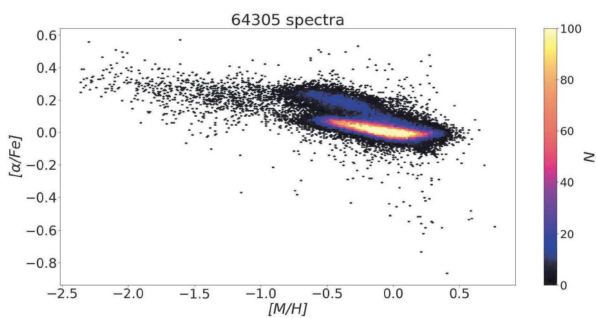

**Fig. 5.** [M/H] vs. [$\alpha$/Fe] distribution for the sample of reference stars, with parameter values obtained from APOGEE DR17 and GALAH DR3, see text for details.

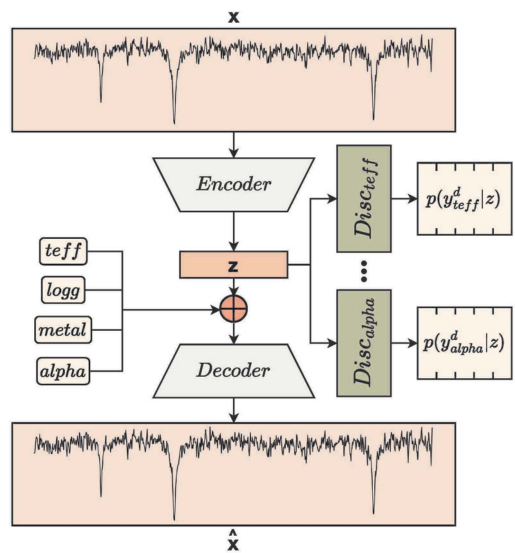

**Fig. 6.** Schematic of the adversarial network for the use case where the value of the astrophysical parameters cannot be provided as input.

### 3.3. Results

The training of the GANs was carried out with 51,444 spectra corresponding to 80% of the set of 64 305 reference stars, reserving the remaining 20% (12 861 spectra) for the inference phase. By using GANDALF, we compute three different latent representations of the spectra, the first one simply using the tool to construct the latent space without any parameter entanglement or disentanglement, and another two by combining the options for disentangling and entangling on parameter pairs. The pairs were selected by affinity, always grouping the physical parameters, $T_{eff}$ and $\log g$ together, and the same for the chemical parameters [M/H] and [$\alpha$/Fe].

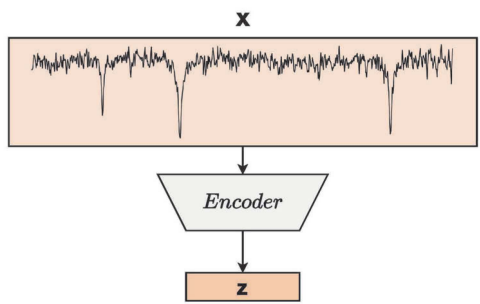

**Fig. 7.** Schematic of the inference phase. Once the adversarial network is trained, the encoder is used to produce the latent space (**z**) of sample spectra with unknown stellar parameters.

Schematically, to test the effectiveness of the approach with disentangling and entangling, four data representations were considered:

 i) RVS spectra. The nominal RVS spectra without any dimensionality reduction.
 ii) Pure latent space. Disable the influence of the 4 discriminators in the error calculation ($\lambda_i = 0$), equivalent to training a regular autoencoder.
 iii) First disentangling/entangling. We disentangle $T_{eff}$ and $\log g$ information from the latent space and we entangle [M/H] and [$\alpha$/Fe].
 iv) Second disentangling/entangling. We entangle $T_{eff}$ and $\log g$ information from the latent space and we disentangle [M/H] and [$\alpha$/Fe].

The t-SNE algorithm can be applied to the spectra and the corresponding latent spaces to assess the level of disentangling and entangling. For clarity, we illustrate both disentangling/entangling scenarios. The t-SNE representation from the first disentangling experiment, which attempts to eliminate Teff and log$g$, are shown on panel (a) of Figure 8, while the second experiment is depicted on panel (b).

Figure 8 also presents the t-SNE diagram for the original spectra on the left column of the subfigures, while the latent representations are depicted on the right ones. It is important to note that the axes of a t-SNE diagram have no physical meaning, simply reflecting the reduced representation produced by the algorithm for each object. The objects in the t-SNE diagrams are colour-coded according to their parameter values, as each figure indicates.

For example, in panel (a), the top two figures show that the latent space has lost information about Teff and log$g$, whereas the bottom figures indicate that the information about [M/H] and [$\alpha$/Fe] has been enhanced. Thus, t-SNE serves as a useful tool to verify the effectiveness of GANDALF in generating latent space representations with an appreciable level of disentangling or reinforcement.

Once the latent representations of the data have been obtained using GANDALF, stellar parameters for the 12 861 stars are extracted using simple ANNs as regressors. After training, the ANNs deliver continuous estimates of the stellar parameters. The architecture is the following: the input is composed of 800 neurons (in the case of the original spectra) or 25 neurons (for the latent representations), 2 hidden layers of 200 and 100 neurons, and the output is composed of 4 neurons to map the 4 stellar parameters. All experiments in this work were conducted using Scikit's (Pedregosa et al. 2011) fully connected networks with backpropagation, maintaining the same structure across all cases.

Tests were conducted to train networks for predicting individual parameters as well as all four parameters simultaneously. Since the results were highly similar, only those for the network predicting all four stellar parameters are presented.

The results are summarised in Table 1, showing the mean, absolute mean, and standard deviation values of the differences to the literature values for the four stellar parameters. While the statistics are generally good for both the pure latent space representation and the original spectra illustrating the effectiveness of ANNs as regressors, it is clear that using latent space representations with disentangling of non-target parameters and enhancement of target parameters significantly improves the outcomes. The accuracy obtained is sufficient to distinguish between dwarf and giant stars, between objects belonging either to the halo or to the thin and thick components of the galactic disc and consequently, to carry out studies of stellar populations. Figures 9 and 10 show the Kiel and [M/H] vs. [$\alpha$/Fe] diagrams obtained with entangling of the relevant parameters ($T_{eff}$ and $\log g$ in the case of the Kiel diagram and [M/H] and [$\alpha$/Fe] for the other diagram) and disentangling of the other pair of parameters.

Regarding computational times, a threefold reduction is achieved when training the regressors with the optimised latent space instead of the original spectra over 1000 iterations on a six-core desktop computer, an Intel Core i7 8700 at 3.20 GHz, and 96 GB of RAM.

## 4. Conclusions

Discovering data representations that facilitate the extraction of useful information and enhance algorithm performance in classification and parameterisation problems is essential in the context of extensive spectroscopic astronomical surveys. This work introduces an encoder-decoder architecture, featuring a modification of the traditional autoencoder with adversarial training. The objective is to disentangle the desired parameters from the rest of the information contained in astronomical spectra within the latent space produced by the encoder. Specifically, we have developed an algorithm for the physicochemical disentanglement of information present in the RVS Gaia spectroscopic survey. By projecting the signal (stellar spectra) into a latent space where contributions from physical properties (Teff and logg) are minimised, we can maximise the information related to chemical properties, thereby improving their extraction. The disentangling of atmospheric chemical parameters [M/H] and [$\alpha$/Fe] also enhances the precision in the derivation of physical parameters.

The level of disentanglement in the latent space is visualised using the t-SNE algorithm. Beyond its initial application, we believe that our methodology can significantly impact Big Data Astronomy, especially given the rise of modern all-sky spectroscopic surveys as our algorithm offers substantial data dimensionality reduction while preserving and even highlighting key information. Additionally, our method can be used to efficiently interpolate and extrapolate samples beyond the input data range, useful for validation or creating new synthetic spectral grids. To implement this, we developed an ad-hoc framework called GANDALF. This framework includes several Python classes for data generation, command line tools for model definition, training, and testing, and a web application to visually demonstrate the algorithm. GANDALF is available to the community as an open-source tool on GitHub.

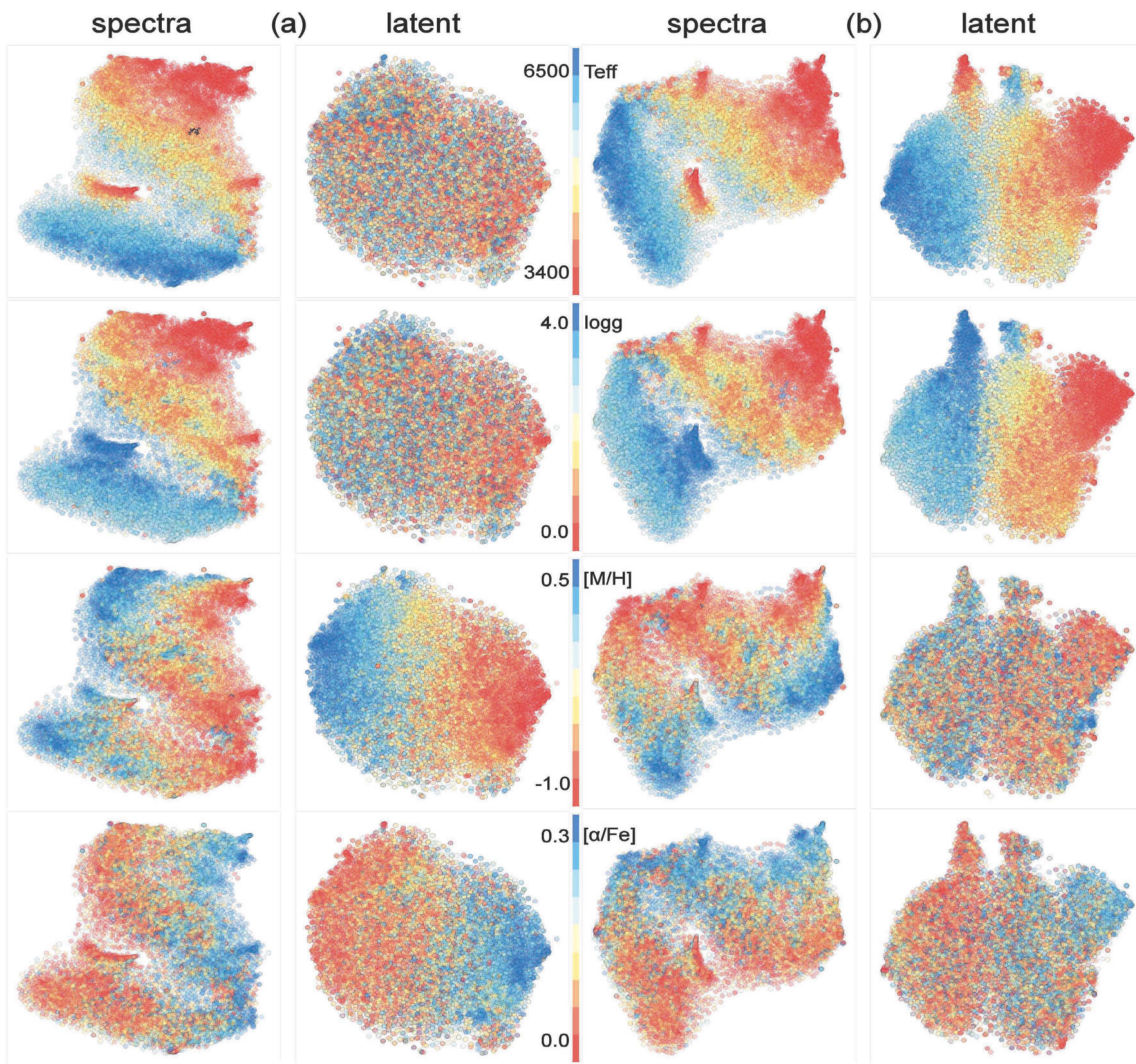

**Fig. 8.** t-SNE diagram obtained for the two disentangling exercises. The column panels (a) refer to the first disentangling exercise, in which we try to eliminate the contribution due to the physical parameters (Teff and logg) in the latent space and reinforce the influence of the chemicals ones ([M/H] and [α/Fe]). The inverse case is shown in column (b). Each panel illustrates the t-SNE result on the original spectrum (subfigures on the left) and the latent space (on the right) corresponding to each disentangling exercise. The figures are colour-coded according to the values of the parameters, as indicated in each figure.

**Table 1.** Summary statistics obtained for parameterisation using different representations of the input space.

|  | RVS Spectra | | | Pure latent | | | First disentangling | | | Second disentangling | | |
|---|---|---|---|---|---|---|---|---|---|---|---|---|
|  | MAE | Mean | $\sigma$ | MAE | Mean | $\sigma$ | MAE | Mean | $\sigma$ | MAE | Mean | $\sigma$ |
| $T_{eff}$(K) | 121 | 58 | 131 | 140 | 6 | 138 | 93 | -7 | 94 | 512 | 17 | 361 |
| log$g$(dex) | 0.231 | 0.017 | 0.277 | 0.282 | -0.012 | 0.340 | 0.141 | -0.021 | 0.160 | 0.868 | -0.027 | 0.522 |
| [M/H](dex) | 0.099 | -0.007 | 0.143 | 0.100 | -0.016 | 0.111 | 0.183 | -0.017 | 0.166 | 0.065 | 0.002 | 0.063 |
| [α/Fe](dex) | 0.039 | -0.001 | 0.053 | 0.041 | 0.006 | 0.043 | 0.051 | 0.003 | 0.049 | 0.031 | -0.004 | 0.033 |

**Notes.** First disentangling means: $T_{eff}$, log$g$ entangled; [M/H], [α/Fe] disentangled. Second disentangling means: $T_{eff}$, log$g$ disentangled; [M/H], [α/Fe] entangled.

*Acknowledgements.* Horizon Europe funded this research [HORIZON-CL4-2023-SPACE-01-71] SPACIOUS project, Grant Agreement no. 101135205, the Spanish Ministry of Science MCIN / AEI / 10.13039 / 501100011033, and the European Union FEDER through the coordinated grant PID2021-122842OB-C22. We also acknowledge support from the Xunta de Galicia and the European Union (FEDER Galicia 2021-2027 Program) Ref. ED431B 2024/21, CITIC ED431G 2023/01, and the European Social Fund - ESF scholarship ED481A2019/155.

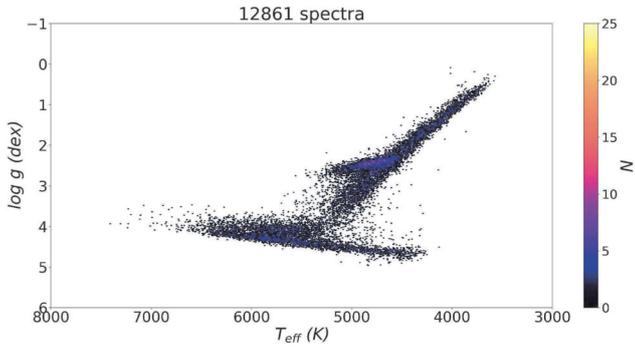

**Fig. 9.** Kiel diagram for test stars obtained after parameterisation with latent spaces using entangling for $T_{eff}$ and $\log g$, and disentangling for [M/H] and [$\alpha$/Fe].

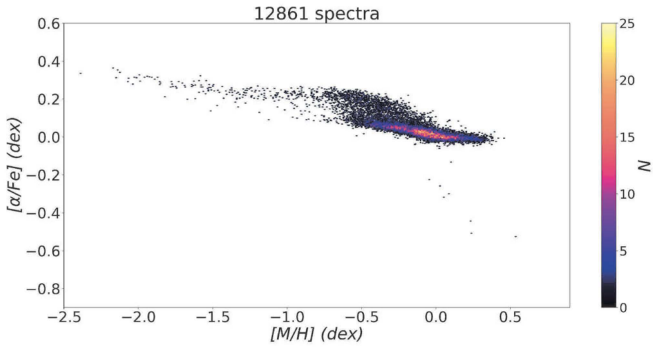

**Fig. 10.** [M/H] vs. [$\alpha$/Fe] diagram for test stars obtained after parameterisation with entangling of [M/H] and [$\alpha$/Fe], and disentangling of $T_{eff}$ and $\log g$.